\documentclass[a4paper]{article}%
\usepackage{times}
\usepackage{amssymb}
\usepackage{graphicx}
\usepackage{amsmath}
\usepackage{epsfig}
\usepackage{amsfonts}%
\usepackage{latexsym}
\begin{document}

\title{Testing of Metric-Field Equations of Gravitation by Binary Pulsar }
\author{L. V. Verozub\footnote{verozub@gravit.kharkov.ua}
 and A. Ye. Kochetov\footnote{kochetov@astron.kharkov.ua}
\\Kharkov National University, Ukraine}
\date{}
\maketitle

\begin{abstract}
Testing of the gravitation equations, proposed by one of the authors
earlier, by a binary pulsar is considered. It has been shown that the 
formulas
for the gravitation radiation of the system resulting from the equations 
do not
contradict the available observations data
\end{abstract}

\section{Introduction}

In paper \cite{Verozub1} gravitation equations which do not lead to a 
physical
singularity in the center of the spherically symmetric field were proposed.
These equations also predict that there can exist stable supermassive compact
configurations of the degeneration Fermi - gas without an events horizon
\cite{Verozub2}. The equations do not contradict classical tests at the
distances from the center which are much larger than the Schwarzschild radius.
In the present paper we find the power of the gravitational-wave radiation
from a close binary system and use the result to find the deceleration of 
the
orbital period of the pulsar PSR1913+16 conditioned by the gravitational
radiation. In this case we deal with a moderately strong gravitation field 
and use
the definition of gravitational energy that follows from the gravitation
equations under consideration.

%%%%%%%%%%%%%%%%%%%%%%%%%%%%%%%%%%%%%%%%%%%%%%%%%%%%%%%%%%%%%%

\section{Equations of Field}

Thirring \cite{Thirring} proposed that gravitation can be described as a
tensor field $\psi_{\alpha\beta}(x)$ of spin two in Pseudo-Euclidean
space-time $E_{4}$ where the Lagrangian action describing the motion of test
particles in a given field is of the form
\begin{equation}
L=-m_{p}c\left[  g_{\alpha\beta}(\psi)\dot{x}^{\alpha}\dot{x}^{\beta}\right]
^{1/2} . \label{Thirring_Lagr}%
\end{equation}
In this equation $g_{\alpha\beta}$ is a tensor function of 
$\psi_{\alpha\beta
}$, $m_{p}$ is the mass of the particle, $c$ is the speed of light and
$\dot{x}^{\alpha}=dx^{\alpha}/dt$.

A theory based on that action must be invariant under the gauge
transformations $\psi_{\alpha\beta}\longrightarrow\bar{\psi}_{\alpha\beta}$
that are a consequence of the existence of \textquotedblright
extra\textquotedblright\ components of the tensor $\psi_{\alpha\beta}$.
Transformations $\psi_{\alpha\beta}\longrightarrow\bar{\psi}_{\alpha\beta}$
give rise to some transformations $g_{\alpha\beta}$ $\longrightarrow$ $\bar
{g}_{\alpha\beta}$. Therefore, the field equations for $g_{\alpha\beta}(x)$
and equations of the motion of the test particle must be invariant under these
transformations of the tensor $g_{\alpha\beta}$. Equations of gravitation that
are invariant with respect to arbitrary gauge transformations were proposed in
\cite{Verozub1}. These equations are of the form
\begin{equation}
B_{\alpha\beta;\gamma}^{\gamma}-B_{\alpha\delta}^{\epsilon}B_{\beta\epsilon
}^{\delta}=0.\label{myeqs}%
\end{equation}
The equations are vacuum bimetric equations for the tensor
\begin{equation}
B_{\alpha\beta}^{\gamma}=\Pi_{\alpha\beta}^{\gamma}-\overset{\circ}{\Pi
}_{\alpha\beta}^{\gamma}.\label{tensB}%
\end{equation}
(Greek indices run from 0 to 3), where
\begin{equation}
\Pi_{\alpha\beta}^{\gamma}=\Gamma_{\alpha\beta}^{\gamma}-(n+1)^{-1}\left[
\delta_{\alpha}^{\gamma}\Gamma_{\epsilon\beta}^{\epsilon}-\delta_{\beta
}^{\gamma}\Gamma_{\epsilon\alpha}^{\epsilon}\right]  ,\label{Thomases}%
\end{equation}%
\begin{equation}
\overset{\circ}{\Pi}_{\alpha\beta}^{\gamma}=\overset{\circ}{\Gamma}%
_{\alpha\beta}^{\gamma}-(n+1)^{-1}\left[  \delta_{\alpha}^{\gamma}%
\overset{\circ}{\Gamma}_{\epsilon\beta}^{\epsilon}-\delta_{\beta}^{\gamma
}\overset{\circ}{\Gamma}_{\epsilon\alpha}^{\epsilon}\right]
,\label{Thomases0}%
\end{equation}
$\overset{\circ}{\Gamma}_{\alpha\beta}^{\gamma}$ are the Christoffel symbols
of the pseudo-Euclidean space-time $E_{4}$ whose fundamental tensor is
$\eta_{\alpha\beta}$, $\Gamma_{\alpha\beta}^{\gamma}$ are the Christoffel
symbols of the Riemannian space-time $V_{4}$ with dimension $n=4$, whose
fundamental tensor is $g_{\alpha\beta}$. The semi-colon in eqs. (\ref{myeqs})
denotes the covariant differentiation in $E_{4}$.

A peculiarity of eqs.(\ref{myeqs}) is that they are invariant under arbitrary
transformations of the tensor $g_{\alpha\beta}$ retaining invariant the
equations of motion of a test particle, i.e. geodesics in $V_{4}$. In other
words, the equations are geodesic-invariant. Thus, the tensor field
$g_{\alpha\beta}$ is defined up to geodesic mappings of space-time $V_{4}$ (in
the analogous way as the potential $A_{\alpha}$ in electrodynamics is
determined up to gauge transformations). A physical sense have only geodesic
invariant values. The simplest object of that kind is the object
$B_{\alpha\beta}^{\gamma}$ which can be named the strength tensor of the
gravitation field. The coordinate system is defined by the used measurement
instruments and is given.

The Christoffel symbols are transformed under the geodetic (i.e. projective)
mappings as follows:
\begin{equation}
\overset{\_}{\Gamma}_{\beta\gamma}^{\alpha}=\Gamma_{\beta\gamma}^{\alpha
}+\varphi_{\beta}\delta_{\gamma}^{\alpha}+\varphi_{\gamma}\delta_{\beta
}^{\alpha} , \label{ChristoffelTransformation}%
\end{equation}
where $\varphi_{\gamma}$ is a vector-function of $x^{\alpha}$. This equation
has a simple interpretation in an (n+1)-dimensional manifold ${\mathcal{M}%
}_{5}$ where the admissible coordinates transformations are of the form%
\begin{equation}
\overset{\_}{x}^{\alpha}=\overset{\_}{x}^{\alpha}(x^{0},x^{1},x^{2},x^{3}) ,
\label{5coordTransform1}%
\end{equation}%
\begin{equation}
\overset{\_}{x}^{4}=x^{4}-\int\varphi_{\alpha}dx^{\alpha} .
\label{5coordTransform2}%
\end{equation}
Namely, eq. (\ref{ChristoffelTransformation})\ can be interpreted as the
transformation of 4-components $\Gamma_{\beta\gamma}^{\alpha}$ of the
connection coefficient $\Gamma_{BC}^{A}$ ($A,B,C=0..4$) in 
${\mathcal{M}}_{5}$
under the transformation (\ref{5coordTransform2}) if the condition
$\Gamma_{4\beta}^{\alpha}=\delta_{\beta}^{\alpha}$ is satisfied.

For this reason we will consider the tensor $g_{\alpha\beta}$ as 4-components
of 5-dimensional tensor
\begin{equation}
g_{AB}=\left(
\begin{array}
[c]{cc}%
g_{\alpha\beta} & g_{\alpha4}\\
g_{4\alpha} & g_{44}%
\end{array}
\right)  . \label{5dimensionMetricTensor}%
\end{equation}
The components $g_{\alpha4}$ are transformed under (\ref{5coordTransform2}) 
as
follows%
\begin{equation}
\overset{\_}{g}_{\alpha\beta}=g_{\alpha\beta}+g_{44}\varphi_{\alpha}%
\varphi_{\beta}+g_{4\alpha}\varphi_{\beta}+g_{4\beta}\varphi_{\alpha} ,
\label{galphabetaTransform}%
\end{equation}%
\begin{equation}
\overset{\_}{g}_{\alpha4}=g_{\alpha4}+g_{44}\varphi_{\alpha} ,
\label{g4alphaTransform}%
\end{equation}%
\begin{equation}
\overset{\_}{g}_{44}=g_{44} . \label{g44Transform}%
\end{equation}
Eqs. (\ref{g4alphaTransform}) coincides with the transformation of the
object$\ \Gamma_{\alpha\beta}^{\beta}$ (or $Q_{\alpha}=\Gamma_{\alpha\beta
}^{\beta}\>-\>\overset{\circ}{\Gamma}_{\alpha\beta}^{\beta}$) under
(\ref{5coordTransform2}), if $g_{44}=n+1$. For this reason, we will assume
that
\begin{equation}
g_{AB}=\left(
\begin{array}
[c]{cc}%
g_{\alpha\beta} & Q_{\alpha}\\
Q_{\alpha} & n+1
\end{array}
\right)  .
\end{equation}

We will assume also that (\ref{galphabetaTransform}) is the transformation 
of
the tensor $g_{\alpha\beta}$ under the geodesic mappings of $V_{4}$. Then
there exists the geodesic-invariant tensor
\begin{equation}
G_{\alpha\beta}=g_{\alpha\beta}-(n+1)^{-1}Q_{\alpha}Q_{\beta}%
,\label{5dimMetricTensorInvatiant}%
\end{equation}
and the geodesic-invariant generalization of the Einstein equations with
matter source are of the form%
\begin{equation}
B_{\alpha\beta;\gamma}^{\gamma}-B_{\alpha\sigma}^{\gamma}B_{\beta\gamma
}^{\sigma}=k\left(  T_{\alpha\beta}-1/2G_{\alpha\beta}T\right)
,\label{MainEquationWithSource}%
\end{equation}
where $k=8\pi G/c^{4}$, $T_{\alpha\beta}$ is the matter energy-momentum
tensor, $T=G^{\alpha\beta}T_{\alpha\beta}$. At the gauge conditions
$Q_{\alpha}=0$ they coincide with the Einstein equations.

Consider now the question about the definition of the energy of gravitational
field in the used theory. Set in eqs. (\ref{MainEquationWithSource})
$T_{\alpha\beta}=\rho c^{2}u_{\alpha}u_{\beta},$ where $\rho$ is the matter
density and $u_{\alpha}$ is the 4-velocity of matter points. At the small
macroscopic velocities of the matter we can set $u_{0}=1$ and $u_{i}=0$. 
Then,
the 00-component of eq (\ref{MainEquationWithSource}) can be written in the
form%
\begin{equation}
B_{00 ; \beta}^{\beta}=\chi(\rho c^{2}+t_{00}) , \label{00-comp_eqs}%
\end{equation}
where $\chi=k/2$, and $t_{00}$ is the 00-component of the tensor
\begin{equation}
t_{\alpha\beta}=\chi^{-1}B_{\alpha\sigma}^{\gamma}B_{\beta\gamma}^{\sigma} .
\label{TensorEnergyGravField}%
\end{equation}
Setting%
\begin{equation}
B_{00}^{\alpha}=c^{-2}\partial U/\partial x^{\alpha}%
\end{equation}
we obtain the equation
\begin{equation}
\Delta U=\chi(\rho c^{2}+t_{00}) . \label{GeneralLaplaceEquation}%
\end{equation}
\label{Generalisational_Poisson_eq} In the absence of the term $t_{00}$ and 
at
the distances from the central masses much larger than the Schwarzschild
radius this equation coincides with the Poisson equation for the Newtonian
gravity potential. For this reason it is natural to expect that in general
case eq. (\ref{GeneralLaplaceEquation}) is the differential equation for a
generalization of the gravitational potential too and the term $t_{00}$, 
which
is in the equation additively with $\rho c^{2},$ is the energy of the
gravitational field.

To verify this assumption let us find the energy of gravitational field of the
point mass $M$ as the following integral in the Pseudo-Euclidean space-time:
\begin{equation}
\mathcal{E}=\int t_{00} dV . \label{energydef}%
\end{equation}
In the Newtonian theory this integral is divergent. In our case we have
\begin{equation}
t_{00}=2\chi^{-1}B_{00}^{1}\;B_{01}^{0} \label{t00}%
\end{equation}
and, therefore, using spherical coordinates, we find%
\begin{equation}
\mathcal{E}=\int t_{00} dV = \frac{1}{8}\frac{r_{g}^{2}c^{4}}{\pi G}J ,
\label{energycalc}%
\end{equation}
where%
\begin{equation}
J=\int\frac{dV}{f^{4}}=\frac{4\pi}{3r_{g}}B(1,1/3)
\end{equation}
and
\begin{equation}
B(z,w)=\int_{0}^{\infty}\frac{t^{z-1}}{(1+t)^{z+w}}dt \label{B-function}%
\end{equation}
is B-function. Using the equality
\begin{equation}
B(z,w)=\frac{\Gamma(z)\Gamma(w)}{\Gamma(z+w)} , \label{BataGamma}%
\end{equation}
where $\Gamma$ is $\Gamma$-function we obtain
\begin{equation}
\mathcal{E}=\frac{r_{g} c^{4}}{2 G}=M c^{2} . \label{energyfinally}%
\end{equation}

We arrive at the conclusion that the energy of the point mass is finite and
the rest energy of the point particle is caused by its gravitational field.
The spacial components of the vector $P_{\alpha}=t_{0\alpha}$ are equal to
zero. Due to these facts we may consider (at least in a weak-field
approximation) the tensor $t_{\alpha\beta}$ as the energy-momentum tensor of
gravitational field.
%%%%%%%%%%%%%%%%%%%%%%%%%%%%%%%%%%%%%%%%%%%%%%%%%%%%%%%%%%%%%%%%%%%

\section{Gravitation radiation of a binary system}

In no gravitation case $g_{\alpha\beta}=\eta_{\alpha\beta}$, where
$\eta_{\alpha\beta}$ is the metric tensor of Pseudo-Euclidean space-time
$E_{4}$. For this reason it is natural to suppose that in a weak-field
approximation
\begin{equation}
g_{\alpha\beta}=\eta_{\alpha\beta}+h_{\alpha\beta} . \label{gapproximation}%
\end{equation}

To simplify eq. (\ref{MainEquationWithSource}) let us choose the following
gauge condition
\begin{equation}
\eta^{\alpha\beta}D_{\alpha\beta}^{\lambda}+\frac{n-1}{n+1}\eta^{\alpha
\lambda}D_{\alpha\beta}^{\beta}=0 , \label{calibration}%
\end{equation}
where $D_{\alpha\beta}^{\gamma}$ is the Christoffel symbols of space-time
$V_{4}$, in which the derivatives are replaced by the covariant ones in 
$E_{4}
$. Equality (\ref{calibration}) is covariant and, therefore, does not imply
restrictions for choosing the coordinate system. Using (\ref{gapproximation})
we obtain to the first order in $h_{\alpha\beta}$:
%\begin{eqnarray}
%B_{\alpha\beta}^{\gamma}=\frac{1}{2}\left(  h_{\alpha;\beta}^{\gamma}%
%+h_{\beta;\alpha}^{\gamma}-h_{\alpha\beta}^{;\gamma}\right) \\
%-\frac{1}%
%{2}\left(  n+1\right)  ^{-1}\left[  \delta_{\alpha}^{\gamma}h_{\sigma;\beta
%}^{\sigma}+\delta_{\beta}^{\gamma}h_{\sigma;\alpha}^{\sigma}\right]
%\label{Bh}% \nonubber
%\end{eqnarray}%
\begin{equation}
B_{\alpha\beta}^{\gamma}=\frac{1}{2}\left(  h_{\alpha; \beta}^{\gamma
}+h_{\beta; \alpha}^{\gamma}-h_{\alpha\beta}^{ ; \gamma}\right)  -\frac{1}%
{2}\left(  n+1\right)  ^{-1}\left(  \delta_{\alpha}^{\gamma}h_{\sigma; \beta
}^{\sigma}+\delta_{\beta}^{\gamma}h_{\sigma; \alpha}^{\sigma}\right)  .
\label{Bh}%
\end{equation}
and the gauge conditions (\ref{calibration}):
\begin{equation}
h_{\lambda; \beta}^{\beta}=\left(  n+1\right)  ^{-1}h_{\sigma; \lambda
}^{\sigma} . \label{hcalibration}%
\end{equation}

Consider a system of the slowly moving bodies of a finite volume where
$T_{\mu\nu}\not =0$ in the vicinity of the coordinates origin, and suppose
that $\left|  T_{00}\right|  \gg\left|  T_{0i}\right|  \gg\left|
T_{ij}\right|  $ (Latin indexes run from 1 to 3). Using an orthogonal system
of coordinates with the metric $\eta_{\mu\nu}=diag(-1,1,1,1)$ let us find the
gravitation radiation in the wave zone.

Equations (\ref{MainEquationWithSource}) now are of the form
\begin{equation}
\Box\psi_{\alpha\beta}=-2kT_{\alpha\beta} , \label{hWaveEquation}%
\end{equation}
where
\begin{equation}
\psi_{\alpha\beta}=h_{\alpha\beta}-1/2\eta_{\alpha\beta}h_{\lambda}^{\lambda}
. \label{define_psi}%
\end{equation}
The general solution of eq. (\ref{hWaveEquation}) for departing waves is of
the form
\begin{equation}
\psi_{\alpha\beta}=\frac{k}{2\pi}\int{\frac{T_{\alpha\beta}\left(  ct-\left|
\mathbf{r}-\mathbf{r}^{\prime}\right|  \>,\mathbf{r}\>^{\prime}\right)
}{\left|  \mathbf{r}-\mathbf{r}\>^{\prime}\right|  }\>d^{\>(3)}x\>^{\prime}},
\label{psiSolution}%
\end{equation}
where $\mathbf{r}\>^{\prime}$ is the radius-vector of an infinitely small
element of the source and $\mathbf{r}$ is the radius vector of the point in
which the field is calculated.

If the origin of the coordinate system coincides with the masses center of the
radiating system, the solution of eq. (\ref{hWaveEquation}) is given by%
\begin{align}
\psi_{00}=const+ o(r^{-3}), \psi_{0i}=o(r^{-2})\\
\psi_{ij}=\frac{k}{4\pi r} \ddot{I}(ct-r,x^{^{\prime}k})+o(r^{-2}),
\end{align}
where
\begin{equation}
I_{ij}=\frac{1}{c^{2}}\int T_{00}x^{\prime}\>^{i}x^{\prime}\>^{j}%
d^{(3)}x^{\prime} , \label{QuadMomentum}%
\end{equation}
the points denote the differentiation with respect to time $t$ and $const$ is
denotes an additional term to the newtonian potential. Therefore, the solution
of eq. (\ref{psiSolution}) for weak gravitational waves can be written in the
form
\begin{equation}
\psi_{0\mu}=0, \label{conditions1}%
\end{equation}%
\begin{equation}
\psi_{ik}=\frac{k}{4\pi r}\ddot{I}_{ik}(\>ct-r,x^{\prime}{}^{k}) .
\label{hSolution}%
\end{equation}

Therefore, the lowest mode of the gravitational radiation is the
quad\-ru\-po\-le one.

Let us find the number of the independent states of the field $\psi
_{\alpha\beta}$ in vacuum. According to (\ref{hWaveEquation}) the functions
$\psi_{\alpha\beta}$ in vacuum obey the equation
\begin{equation}
\Box\psi_{\alpha\beta}=0\; \label{PsiEquation}%
\end{equation}
and gauge conditions (\ref{hcalibration}), which give four constraint
equations. Equations (\ref{conditions1}) also gives four constraints. Taking
also into account that for weak waves $\psi_{0i}=0$, we have only two
independent components of $\psi_{\alpha\beta}$. It follows also from
(\ref{hcalibration}) and (\ref{conditions1}) that the trace of the tensor
$\psi_{\alpha\beta}$ is equal to zero. Thus,
\begin{equation}
\psi_{0\mu}=0,\;\psi_{ij\>,\>j}=0,\;\psi_{kk}=0 , \label{psiConditions}%
\end{equation}
where the coma denotes the partial derivative. Therefore, the plane
gravitational wave is a transversal with the zero trace.

Now let us find the flow of energy of the gravitational radiation. Starting
from eq.(\ref{TensorEnergyGravField}), we find that the flow of  energy
carried by the wave in the radial direction is given by
\begin{equation}
-ct_{0k}=\frac{n_{k}}{4kc}\langle\dot{\psi}_{ij}\dot{\psi}_{ij}\rangle
,\label{flow}%
\end{equation}
where $n^{k}=x^{k}/r\>$ and the symbols $\langle\;\rangle$ denote the period
$T$-averaging of the a function $f(t)$ by the formula
\begin{equation}
\left\langle f\left(  t\right)  \right\rangle =
\frac{1}{T}\int_{0}^{T}f\left(
t\right)  dt.\label{average}%
\end{equation}
We can also satisfy condition (\ref{psiConditions}) by projecting the tensor
$\psi_{\alpha\beta}$ to the plane perpendicular to the vector $n^{i}$ with 
the
help of the operator
\begin{equation}
P_{ij}=\delta_{ij}-n_{i}n_{j}.\label{projection}%
\end{equation}
As a result, the transversal-traceless part of the tensor 
$\psi_{\alpha\beta}$
is
\begin{equation}
\psi_{ij}^{TT}=P_{ir}P_{js}\psi_{rs}-\frac{1}{2}P_{ij}\left(  P_{rs}\psi
_{rs}\right)  ,\label{psiTT1}%
\end{equation}

It should also be noted that the transversal-traceless part of the
quad\-ru\-po\-le momentum $I_{ij}$, in (\ref{hSolution}), and the magnitude
$J_{ij}=I_{ij}-1/3\delta_{ij}I_{kk}$ are the same, which can be verified
immediately. For this reason
\begin{equation}
\psi_{ij}=\frac{k}{4\pi r}\ddot{J}_{ij} . \label{psiTT2}%
\end{equation}
Finally, we obtain
\begin{equation}
-ct_{0r}=\frac{n_{r}k}{64\pi^{2}r^{2}c}\langle\overset{...}{J}_{ij}%
\overset{...}{J}_{ij}-2\overset{...}{J}_{ik}\overset{...}{J}_{kj}n_{i}%
n_{j}+\frac{1}{2}\overset{...}{J}_{ij}\overset{...}{J}_{kl}n_{i}n_{j}%
n_{k}n_{l}\rangle, \label{FluxOfEnergy}%
\end{equation}

The averaging time rate of the energy flow change can be found by integration
of eq.(\ref{FluxOfEnergy}) over the sphere of the radius $r$:
\begin{equation}
-\left\langle \frac{dE}{dt}\right\rangle =\int ct_{0r}n_{r}r^{2}\sin\theta
d\theta d\varphi=\frac{G}{5c^{5}}\langle\overset{...}{J}_{ij}\overset{...}%
{J}_{ij}\rangle, \label{Power}%
\end{equation}
were $E$ is the total mechanical energy of the radiating system. This formula
completely coincides with the one in general relativity.

Consider a system of two gravitationally bound point masses $m_{1}$ and
$m_{2}$. Assuming that the distance between the bodies is much larger than the
Schwarzschild radiuses we can use the Newtonian mechanics. Since $E=-G\mu
M/(2a)$, where $M=m_{1}+m_{2}$, $\mu=m_{1}m_{2}/M$, $a$ is the large semiaxes
of the orbit and the orbital period is $P_{b}=2\pi a^{3/2}(MG)^{-1/2}$, the
relative velocity of increasing of the orbital period is given by
\begin{equation}
\frac{dP_{b}/dt}{P_{b}}=\frac{3}{2}\frac{da/dt}{a}=-\frac{3}{2}\frac{dE/dt}{E}
. \label{Pdot}%
\end{equation}
It gives
\begin{equation}
\frac{dP_{b}/dt}{P_{b}}=\frac{192\pi}{5c^{5}}\>\left(  2\pi G\right)
^{5/3}P_{b}^{8/3}\>f(e)m_{1}m_{2}M^{-1/3}\;, \label{dP/dt}%
\end{equation}
where $f(e)=\left(  1-e^{2}\right)  ^{-7/2}\left(  1+73/24\>e^{2}%
+37/96\>e^{4}\right)  $. It should be noted that in the above calculations 
the
shift of the periastron is not taken into account since at the used one
period-averaged the shift is very small.
%%%%%%%%%%%%%%%%%%%%%%%%%%%%%%%%%%%%%%%%%%%%%%%%%%%%%%%%%%%%%%%%%%

\section{Application to PSR1913+16}

Let us use the above results for an analysis of the binary pulsar PSR 1913+16
in the framework of the Blandford-Teukolsky model \cite{BlandfordTeukolsky}.
Namely, we take into account the dependence of the phase of the detected
radiation on the following observable parameters that are the functions of 
the
unknown masses $m_{1}$ and $m_{2}$.

1. Increasing the orbital period because of the gravitational radiation
(\ref{dP/dt}).

2. The rate of the periastron shift of the pulsar orbit which according to
\cite{Verozub2} is given by
\begin{equation}
\dot{\omega}=\frac{6\pi GM}{a\left(  1-e^{2}\right)  c^{2}}+\frac{\pi
G^{2}M^{2}}{2a^{2}\>\left(  1-e^{2}\right)  ^{2}\>c^{4}}\>f_{1}\>(e) ,
\label{domega}%
\end{equation}
where $\ f_{1}\>(e)=\left(  54+16e^{2}-e^{4}\right)  $. The second term for
the considered system is about $0.01\%$ of the total value of $\dot{\omega}$
and, therefore must be taken into account.

3. The magnitude%
\begin{equation}
\gamma=\frac{t_{1}-t_{2}}{\sin(\widetilde{E})} , \label{GammaParameter}%
\end{equation}
where $t_{1}$ is the moment of the radiation of the pulse, measured in the
inertial frame of reference of the distant observer, $t_{2}$ is the same
moment, measured in the proper frame of reference of the pulsar,
$\widetilde{E}$ is the eccentric anomaly.

Supposing that 4-vector $k^{a}=\{\omega/c,\mathbf{k}\}$ of a photon, the
motion of which is described by the Lagrangian (\ref{Thirring_Lagr}),
satisfies the equality $g_{\alpha\beta}k^{\alpha}k^{\beta}=0$, we obtain
approximately $k^{0}=\omega/c\sqrt{g_{00}}$, where $\omega$ is the frequency
of the electromagnetic wave and $g_{00}$ is given by \cite{Verozub1}
\begin{equation}
g_{00}=1-r_{g}/(r_{g}^{3}+r^{3})^{1/3} .
\end{equation}

Let $v_{p}$ be the pulsar velocity relative to the masses center of the
system. Then we obtain the following relation between the measured 
frequencies
of the signal $\omega_{0}$ and $\omega$ in the frames of the observer and
pulsar%
\begin{equation}
\frac{\omega}{\sqrt{1-r_{g}/f_{c}}}=\omega_{0}\frac{1-(v_{p}/c)\cos
(\vartheta)}{\sqrt{1-v_{p}^{2}/c^{2}}} , \label{frequenciesRelation}%
\end{equation}
where $\vartheta$ is the angle between the wave direction and the direction 
of
the motion of the source radiation, $f_{c}=(r^{3}+r_{g})^{1/3},$ $r_{g}$ is
the gravitational radius of the pulsar companion, $r$ is the distance between
the pulsar and its companion. From (\ref{frequenciesRelation}) we obtain,
setting $\vartheta=\pi/2$, the relation between time intervals $d\tau$ and
$dt$ in the reference frames of the pulsar and observer
\begin{equation}
d\tau=\left(  1-\frac{r_{g}}{2r}-\frac{v_{1}^{2}}{2c^{2}}\right)  dt ,
\label{relationBetweenTimes}%
\end{equation}
which with the used accuracy coincides with the same relation in general 
relativity.

The Shapiro effect is too small (the same as in general relativity) and is 
not
taken into account in the model under consideration.

Proceeding from these results, we obtain in an analogy with
\cite{BlandfordTeukolsky}
\begin{equation}
\gamma=\frac{GP_{b}em_{2}^{2}(m_{1}+2m_{2})}{2\pi a c^{2}(m_{1}+m_{2})^{2}} .
\label{ParameterGammaPulsar}%
\end{equation}

Fig. 1 shows that in the plane $m_{1}m_{2}$ the curves $\overset{\cdot}{P}%
_{b}=\overset{\cdot}{P}_{b}(m_{1,}m_{2})$, $\overset{\cdot}{\omega}%
=\overset{\cdot}{\omega}(m_{1,}m_{2})$, $\gamma=\gamma(m_{1,}m_{2})$ 
intersect
in one point that means that the theory does not contradict the observation
data. The value of the full system mass $M$, resulting from the measured 
value
of $P_{b}$ and $e$ by eq.(\ref{domega}) is equal to $(2.82845\pm
0.00004)M_{\odot}$ and the masses of the pulsar and its companion, found by
(\ref{ParameterGammaPulsar}), are equal to $(1.441\pm0.003)M_{\odot}$ and
$(1.387\pm0.003)M_{\odot}$, respectively. These results differ very little
from the ones resulting from general relativity \cite{DamourTaylor92a}
($(1.442\pm0.003)M_{\odot}$ and $(1.386\pm0.003)M_{\odot}$). Due to a
kinematic effect in our Galaxy \cite{DamourTaylor92b} the small correction
$(-0.017\pm0.005)\cdot10^{-12}$ must be added to the found theoretical value
of $\overset{\cdot}{P}_{b}=(-2.40249\pm0.00029)\cdot10^{-12}$. Taking into
account this correction we obtain that the ratio of the observational value 
of
$\overset{\cdot}{P}_{b}$ to the found theoretically is equal to $1.0023\pm
0.0047$.

\vspace{0.5cm}

\centerline{\hbox{ \psfig{file=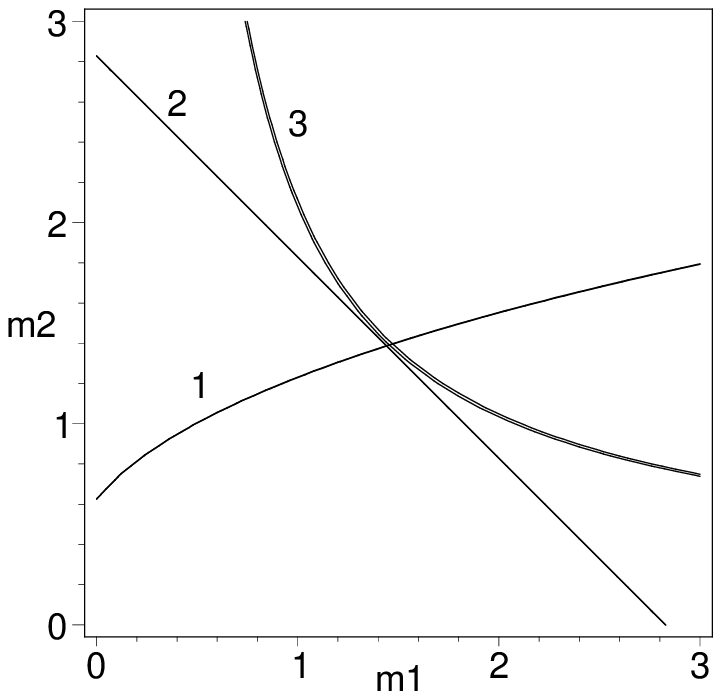,height=5cm,width=5cm}}}

\vspace{0.5cm}

Fig.1 The plots of the functions $\gamma=\gamma(m_{1},m_{2})$ (curve 1),
$\omega=\omega(m_{1},m_{2})$ (curve 2), $\dot P_{b}=\dot P_{b}(m_{1},m_{2})$
(curve 3).


\begin{thebibliography}{9}                                                                                                %


\bibitem {Verozub1}{\small Verozub L.V., Phys. Lett. A, 156 (1991) 404 }

\bibitem {Verozub2}{\small Verozub L.V., Astron. Nachr. 317, 2 (1996) 107 }

\bibitem {Thirring}{\small Thirring W., Ann. Phys. 16 (1961) 96. }

\bibitem {BlandfordTeukolsky}{\small Blandford R. and Teukolsky S. A.,
Astophys. J. 205 (1975) 580. }

\bibitem {DamourTaylor92a}{\small Taylor J. H., Wolszczan A., Damour T. and
Weisberg J. M., Nature, 355 (1992) 132. }

\bibitem {DamourTaylor92b}{\small Damour T. and Taylor J. H., Phys. Rev. D, 
45
(1992) 1840 }
\end{thebibliography}
\end{document}